\documentclass[aps,prl,twocolumn,groupedaddress]{revtex4}

\usepackage{amsmath2000}
\usepackage{amsfonts,bm,graphicx}

\newcommand{\Gfs}{Green functions }

\newcommand{\C}{\v{C}\'apek }
\newcommand{\eps}{\epsilon}

\begin{document}

\title{On apparent breaking the second law of thermodynamics in quantum transport studies}

\author{Tom\'a\v s Novotn\'y}
\email{tno@mag.mff.cuni.cz}

\affiliation{Department of Electronic Structures,
     Faculty of Mathematics and Physics, Charles University,
     Ke Karlovu 5, 121 16 Prague, Czech Republic}

\date{\today}

\begin{abstract}
We consider a model  for stationary electronic transport through a
one-dimensional chain of two leads attached to a perturbed central
region (quantum dot) in the regime where the theory proposed
recently by \C for a similar model of phonon transport predicts
the striking phenomenon of a permanent current between the leads.
This result based on a rigorous but asymptotic Davies theory is at
variance with the zero current yielded by direct transport
calculations which can be carried out in the present model. We
find the permanent current to be within the error of the
asymptotic expansion for finite couplings, and identify cancelling
terms of the same order.
\end{abstract}

\pacs{05.30.-d, 05.70.-a, 44.90.+c}

\maketitle

In recent years, the validity of the second law of thermodynamics
has been questioned in several models
\cite{cap-pre-98,she-pre-98,nikulov,all-nie-prl-00,weiss,capek}.
In most of them the phenomenon of breaking the law was claimed to
stem from the genuine quantum behaviour of the system considered,
e.g.\ quantum entanglement between the system and bath. Several
models support the idea that the quantum correlation features
could be responsible for breaking the second law.

In spite of the seemingly academic nature of the models, had their
predictions turned out to be true, it could have far reaching
consequences in the practice. For example, in the rapidly
developing field of nanotechnologies, the regimes of validity of
the suggested models could be achieved with the present or soon
incoming technology so that the predicted effects, if confirmed,
could be used to supply power for the nano-machines
\cite{nikulov}. Thus, thorough experimental as well as further
theoretical examination of this kind of models is a desirable
task.

In this study, we consider a modified version of the model
molecular 'demon' operating between two reservoirs recently
proposed by \C \cite{capek}. Namely, we study the electronic
transport in a one-dimensional tight-binding chain the central
part of which is considered as 'system' and the rest of the chain
forms two 'reservoirs'. While in a general fermionic case there
might arise questions about the applicability of the standard
projection techniques which assume separability of the system and
bath Hilbert spaces this problem is harmless in our model since an
exact mapping onto an excitonic chain using Jordan-Wigner
transformation \cite{tsvelik} can be performed.

Then the standard techniques are fully justified and using the
method of the modified Davies weak coupling theory from
\cite{cap-bar-pha-01} and repeating the reasoning of \cite{capek}
we come to the result that a permanent stationary current
(particle flow) between the two asymptotic parts of the chain is
possible thus violating the second law. However, the present model
can be solved explicitly in the full range of parameters, i.e.\
without the need to resort to any limit. The exact result
contradicts the one obtained by the modified Davies weak coupling
theory by \v{C}\'apek. Now, let us consider the model in detail.

The Hamiltonian $H=H_S+H_B+H_T$ reads
\begin{align}\label{hamiltonian}
  H_S &= \eps_1 a^{\dag}_1 a_1 + \eps_2 a^{\dag}_2 a_2 +
  J\,(a^{\dag}_1 a_2 + a^{\dag}_2 a_1) \ ,\\
  H_B &= \sum_{\alpha=1,2}\sum_{n=1}^{\infty}
  -\frac{\Delta}{2}(c^{\dag}_{n+1\alpha}c_{n\alpha}
  + c^{\dag}_{n\alpha}c_{n+1\alpha}) \notag \\
   &=\sum_{\bm{k},\alpha=1,2} \eps_{\bm{k}\alpha}\,
  c^{\dag}_{\bm{k}\alpha}c_{\bm{k}\alpha} \ ,\\
  H_T &= \sum_{\alpha=1,2}  V_{\alpha} (c^{\dag}_{1\alpha}a_{\alpha}
  + a^{\dag}_{\alpha} c_{1\alpha}) \notag\\
  &=\sum_{\bm{k},\alpha=1,2} (V_{\bm{k}\alpha}
  c^{\dag}_{\bm{k}\alpha}a_{\alpha} + V^*_{\bm{k}\alpha}
  a^{\dag}_{\alpha}c_{\bm{k}\alpha}) \ ,
\end{align}
where $H_S$ denotes the part corresponding to the system, i.e.\
the quantum dot, $H_B$ describes the contacts (baths) and $H_T$
the tunneling between the respective contact and level of the dot.
The second form of $H_B, H_T$ is expressed in the eigenstates of
the baths labeled by $\bm{k}$. All the creation and annihilation
operators
$a^{\dag}_{\alpha},\ a_{\alpha},\ c^{\dag}_{n\alpha},\ c_{n\alpha},
(\text{or } c^{\dag}_{\bm{k}\alpha},\ c_{\bm{k}\alpha})$ satisfy
the anticommutation relations for the Fermi operators among
themselves. It is assumed that $J, V_{\alpha}, \text{ and }
\Delta$ are real numbers.

In order to fully specify the problem we have to supply the
initial conditions. Let us assume that before the initial time
$t_0$ the system is decoupled from the contacts (i.e.\ $H_T$ is
effectively equal to zero) and a particular contact is in the
thermal equilibrium given by the values of the temperature
$T_{\alpha}$ and electrochemical potential $\mu_{\alpha}$.
The corresponding initial density matrix of the whole system is
thus separable
$\rho_{S+B}(t_0)=\rho_S(t_0)\otimes\rho_B^{\text{can}}$. At $t=t_0$ we
switch on the tunneling interaction between the dot and the
contacts and study the time evolution of the whole complex system
dot plus baths.

The particle current into bath $\alpha,\ I_{\alpha}(t)$, is given
by the time derivative of the particle number
$N_{\alpha}=\sum_{\bm{k}}c^{\dag}_{\bm{k}\alpha}c_{\bm{k}\alpha}
=\sum_{n=1}^{\infty}c^{\dag}_{n\alpha}c_{n\alpha}$ as ($\hbar=1$
in the whole paper)
\begin{equation}
\begin{split}
  I_{\alpha}(t) &= \langle\dot{N}_{\alpha}\rangle(t)
  =-i\langle[N_{\alpha},H]\rangle(t) \\
  &= -i \sum_{\bm{k}}\bigl( V_{\bm{k}\alpha}\langle c^{\dag}_{\bm{k}\alpha}
  a_{\alpha}\rangle(t) - V^*_{\bm{k}\alpha}\langle
  a^{\dag}_{\alpha}c_{\bm{k}\alpha}\rangle(t)\bigr) \\
  &= -i  V_{\alpha}\bigl(\langle c^{\dag}_{1\alpha}a_{\alpha}\rangle(t)
  - \langle a^{\dag}_{\alpha}c_{1\alpha}\rangle(t)\bigr) \ ,
\end{split}
\end{equation}
where the mean value of an arbitrary operator $O$ is calculated by
\begin{equation}
\begin{split}
  \langle O \rangle(t) &=
  \mathrm{Tr}_{S+B}\bigl(\rho_{S+B}(t_0)\,O(t)\bigr) =
  \mathrm{Tr}_{S+B}\bigl(\rho_{S+B}(t)\,O\bigr) \\
  & =
  \mathrm{Tr}_{S+B}\bigl(\rho_{S+B}(t_0)\,e^{iH(t-t_0)}\,O\,e^{-iH(t-t_0)}\bigr)
  \ .
\end{split}
\end{equation}
The particle current between any two adjacent sites is given by
the analogy of the above formula with the creation and
annihilation operators of the two sites
\cite{car-com-noz-stj-jpc-71}.

In the stationary state the current does not depend on time and
neither on the position of the sites between which it is evaluated
(current is conserved and his divergence in the stationary state
is thus zero). Therefore,
\begin{equation}\label{current}
  I^{\text{stat}}_1 = -i J \bigl(\langle a^{\dag}_1 a_2\rangle
  - \langle a^{\dag}_2 a_1\rangle \bigr) = -I_2^{\text{stat}}\ .
\end{equation}

The solution of the model within the formalism of nonequilibrium
\Gfs is well known, see e.g.\
\cite{car-com-noz-stj-jpc-71,jau-win-mei-prb-94,jauho}. Therefore,
we only briefly summarize the results here. If we define the
one-particle density matrix of the dot by
\begin{equation}\label{one-particle}
  \sigma_{\alpha\beta}(t) = \langle a^{\dag}_{\beta}(t)a_{\alpha}(t)\rangle
   = -i G^{<}_{\alpha\beta}(t,t)
\end{equation}
we get for the present model the result
\begin{equation}\label{density}
\begin{split}
    \sigma(t) &= \mathbf{G}^R(t-t_0)\cdot \sigma(t_0)
                          \cdot\mathbf{G}^A(t_0-t) \\
                        &+ \int_{t_0}^t dt_1\,dt_2
                          \mathbf{G}^R(t-t_1)\cdot -i\mathbf{\Sigma}^{<}(t_1-t_2)
                          \cdot\mathbf{G}^A(t_2-t)
\end{split}
\end{equation}
where the retarded and advanced \Gfs and the lesser selfenergy
$\mathbf{G}^{R,A}(t),\mathbf{\Sigma}^{<}(t)$ are given as inverse
Fourier transforms of
\begin{align}
  \mathbf{G}^{R,A}(\eps) &=
  \begin{pmatrix}
     \eps - \eps_1 - \Sigma^{R,A}_1(\eps) & -J \\
     -J & \eps - \eps_2 - \Sigma^{R,A}_2(\eps)
  \end{pmatrix}^{-1} \\
  \mathbf{\Sigma}^{<}(\eps) &=
  \begin{pmatrix}
     i\Gamma_1(\eps)f_1(\eps) & 0 \\
     0 & i\Gamma_2(\eps)f_2(\eps)
  \end{pmatrix}
\end{align}
with
\begin{align}
  \Sigma^{R,A}_{\alpha}(\eps) &= \Lambda_{\alpha}(\eps)\mp \frac{i}{2}
  \Gamma_{\alpha}(\eps) = \sum_{\bm{k}}\frac{|V_{\bm{k}\alpha}|^2}
  {\eps-\eps_{\bm{k}\alpha}\pm i\eta} \notag \\
  &= V_{\alpha}^2 g^{R,A}_{\alpha}(1,1;\,\eps) \\
  f_{\alpha}(\eps) &=\frac{1}{e^{\beta_{\alpha}(\eps-\mu_{\alpha})}+1}\ .
\end{align}
The \Gfs $g^{R,A}_{\alpha}(1,1;\,\eps)$ are given by the matrix
elements of the resolvents of the respective reservoirs and the
parameters $\beta_{\alpha},\ \mu_{\alpha}$ denote their inverse
temperature and electrochemical potential in the initial state.

In the limit $t_0\to -\infty$ which corresponds to the stationary
state the first (initial conditions) term in \eqref{density}
vanishes and the second one can be simplified by employing the
Fourier transform yielding the final expression for the current
\begin{equation}\label{exact}
  I_1 = \int_{-\infty}^{\infty}\frac{d\eps}{2\pi}\,\Gamma_1(\eps)\Gamma_2(\eps)
  G^R_{12}(\eps)G^A_{21}(\eps)\bigl(f_2(\eps)-f_1(\eps)\bigr) \ .
\end{equation}
One can immediately see that for the two bath being originally at
the same temperature and electrochemical potential there is no
stationary current between them. Also for different initial
temperatures and/or electrochemical potentials the formula
predicts correctly the stationary flow in the direction against
the temperature and/or concentration gradient. This result is {\em
exact}, i.e.\ it is valid for any values of the parameters in the
Hamiltonian and, therefore, in the limit assumed by the modified
Davies theory too.

Now, we want to express the current in terms of quantities
employed in the modified Davies weak coupling formalism by
\v{C}\'apek. First, it should be noted that the presently
considered model of a {\em linear} chain with at maximum {\em
nearest-neighbor interaction} populated by fermions can be exactly
mapped by the Jordan-Wigner transformation on the linear chain of
Frenkel excitons which satisfy the Pauli commutation relations,
i.e.\ anticommute on-site and commute at different sites, with the
same form of the Hamiltonian expressed in terms of exciton
operators, see \cite{tsvelik}. Thus, the projection formalism used
by \C can be safely used since no sign problem due to fermions
appears. In the following, we solve by the projection method the
corresponding exciton model but the results for the level
occupations and the current are exactly equal to those of the
fermion model.

Within the projection formalism only the reduced density matrix of
the states of the dot is considered. The basis of the states in
the Hilbert space of the dot corresponds to two excitonic levels
and, thus, may be chosen to be
$\mathcal{H}_S= \{|0\rangle\equiv|\mathrm{vac}\rangle,|1\rangle\equiv a^{\dag}_1|0\rangle,
|2\rangle\equiv a^{\dag}_2|0\rangle,|3\rangle\equiv a^{\dag}_2
a^{\dag}_1 |0\rangle\}$. The reduced density matrix
$\rho(t)$ with the matrix elements $\rho_{ij}(t) =
\langle i |\rho(t)| j\rangle$ is given by
\begin{equation}\label{reduced}
\begin{split}
  \rho(t) &= \mathrm{Tr}_B\bigl(\rho_{S+B}(t)\bigr)
  = \mathrm{Tr}_B\bigl(e^{-i\mathcal{L}(t-t_0)}\rho_{S+B}(t_0)\bigr) \\
  &= \mathrm{Tr}_B\bigl(e^{-iH(t-t_0)}
  \rho_{S+B}(t_0) e^{iH(t-t_0)}\bigr)
\end{split}
\end{equation}
with $\mathcal{L}\,\,\bullet=[H,\,\bullet]$ being the Liouville
superoperator of the whole system. The relation between the
one-fermion density matrix \eqref{one-particle} and the above
introduced exciton reduced density matrix \eqref{reduced} is as
follows
\begin{equation}
\begin{split}
    \sigma_{11}(t) &= \rho_{11}(t)+\rho_{33}(t) \\
    \sigma_{22}(t) &= \rho_{22}(t)+\rho_{33}(t) \\
    \sigma_{12}(t) &= \rho_{12}(t) = \rho_{21}^*(t) =
    \sigma_{21}^*(t) \ .
\end{split}
\end{equation}

The modified Davies approach of \cite{cap-bar-pha-01} is valid in
the limit $\lambda\to 0$ when the following scaling is performed
$V_{\alpha}\to\lambda V_{\alpha}\text{ and } J\to\lambda^2 J$.
In this limit the reduced density matrix obeys the evolution
equation
\begin{equation}\label{EOM}
  \frac{d\rho(t)}{dt} = -i\bigl(\mathcal{L}_S +
  \langle\mathcal{L}_T\rangle + i\mathcal{K}\bigr)\,\rho(t)
\end{equation}
with the superoperator $\mathcal{L}_S$ corresponding to the dot
Hamiltonian $H_S$ and
\begin{align}
  \langle\mathcal{L}_T\rangle\,\,\bullet &= \mathrm{Tr}_B\bigl([H_T,\,
  \bullet\otimes\rho_B^\mathrm{can}]\bigr) \\
  \mathcal{K}\,\,\bullet &=
  \int_0^{\infty}d\tau\Bigl\{\mathrm{Tr}_B\bigl((-i\mathcal{L}_T(\tau))(-i\mathcal{L}_T)
  (\bullet\otimes\rho_B^\mathrm{can})\bigr) \notag \\
  &\qquad\qquad\quad - \langle\mathcal{L}_T(\tau)\rangle
  \langle\mathcal{L}_T\rangle\,\bullet\Bigr\}
\end{align}
where $\mathcal{L}_T(\tau) = e^{i\mathcal{L}_0\tau} \mathcal{L}_T
e^{-i\mathcal{L}_0\tau}$ corresponds to the tunneling Hamiltonian
$H_T(\tau)=e^{iH_0\tau} H_T e^{-iH_0\tau}$ in a modified interaction
picture with $H_0=H_B+H_S\bigl|_{J=0}$. It is this modification of
the time dependence of the interaction superoperator entering the
kernel $\mathcal{K}$ which distinguishes the \v{C}\'apek's
approach from the original Davies weak coupling theory. To be
explicit, the above equation for the elements we are interested in
reads
\begin{widetext}
\begin{equation}\label{matrix}
    \frac{d}{dt}
    \begin{pmatrix}
  \rho_{00} \\ \rho_{11} \\ \rho_{22} \\ \rho_{33} \\
  \text{Re}\rho_{12} \\ \text{Im}\rho_{12}
  \end{pmatrix}
  =
  \begin{pmatrix}
    -(\Gamma_{1\uparrow}+\Gamma_{2\uparrow}) &
    \Gamma_{1\downarrow} & \Gamma_{2\downarrow} & 0 & 0 & 0 \\
    \Gamma_{1\uparrow} &
    -(\Gamma_{1\downarrow}+\Gamma_{2\uparrow}) & 0 &
    \Gamma_{2\downarrow} & 0 & -2\lambda^2 J \\
    \Gamma_{2\uparrow} & 0 &
    -(\Gamma_{1\uparrow}+\Gamma_{2\downarrow}) & \Gamma_{1\downarrow} & 0 & 2\lambda^2 J \\
    0 & \Gamma_{2\uparrow} & \Gamma_{1\uparrow} &
    -(\Gamma_{1\downarrow}+\Gamma_{2\downarrow}) & 0 & 0 \\
    0 & 0 & 0 & 0 & -\frac{1}{2}(\Gamma_1+\Gamma_2) &
    \eps_1'-\eps_2' \\
    0 & \lambda^2 J & -\lambda^2 J & 0 & \eps_2'-\eps_1' & -\frac{1}{2}(\Gamma_1+\Gamma_2)
  \end{pmatrix}
  \cdot
  \begin{pmatrix}
  \rho_{00} \\ \rho_{11} \\ \rho_{22} \\ \rho_{33} \\
  \text{Re}\rho_{12} \\ \text{Im}\rho_{12}
  \end{pmatrix}
\end{equation}
\end{widetext}
where
\begin{equation}\label{rates}
\begin{split}
  \Gamma_{\alpha\uparrow} &= 2\pi\lambda^2\sum_{\bm{k}} |V_{\bm{k}\alpha}|^2
  f_{\alpha}(\eps_{\bm{k}\alpha})\,\delta(\eps_{\alpha}-\eps_{\bm{k}\alpha}) \\
  \Gamma_{\alpha\downarrow} &= 2\pi\lambda^2\sum_{\bm{k}} |V_{\bm{k}\alpha}|^2
  \bigl(1-f_{\alpha}(\eps_{\bm{k}\alpha})\bigr)\,\delta(\eps_{\alpha}-\eps_{\bm{k}\alpha})
  \\
  \Gamma_{\alpha} &= \Gamma_{\alpha\uparrow}+\Gamma_{\alpha\downarrow} =
  \lambda^2\Gamma_{\alpha}(\eps_{\alpha}) \\
  \eps_{\alpha}' &= \eps_{\alpha} +
  \lambda^2\sum_{\bm{k}}\frac{|V_{\bm{k}\alpha}|^2}{\eps_{\alpha}-\eps_{\bm{k}\alpha}}
  = \eps_{\alpha} + \lambda^2\Lambda_{\alpha}(\eps_{\alpha}) \ .
\end{split}
\end{equation}

The current from the site $2$ to $1$ which is in the {\em
stationary} state equal to the current into the first reservoir is
given by
$I_{12}(t) = -2\lambda^2 J\,\text{Im}\rho_{12}(t)$ and from the
above equation can be found to be
\begin{equation}\label{result}
  I^{\text{stat}}_1 = I_{12}^{\text{stat}} = \mathcal{T}_{12}(\rho_{22}-\rho_{11}) =
  \mathcal{T}_{12}(\sigma_{22}-\sigma_{11})
\end{equation}
with the transmission coefficient $\mathcal{T}_{12}$
\begin{equation}\label{transmission}
\begin{split}
  \mathcal{T}_{12} &=2\pi\lambda^4\, J^2\frac{1}{2\pi}\frac{\Gamma_1+\Gamma_2}
  {\bigl(\eps_1'-\eps_2'\bigr)^2+\bigl(\frac{1}{2}(\Gamma_1+\Gamma_2)\bigr)^2}
  \\ &= 2\pi\lambda^4\,J^2\int_{-\infty}^{\infty}d\eps\frac{1}{2\pi}\frac{\Gamma_1}
  {\bigl(\eps-\eps_1'\bigr)^2+\bigl(\frac{1}{2}\Gamma_1\bigr)^2} \\
  & \qquad\qquad\qquad\quad\cdot
  \frac{1}{2\pi}\frac{\Gamma_2}
  {\bigl(\eps-\eps_2'\bigr)^2+\bigl(\frac{1}{2}\Gamma_2\bigr)^2} \ .
\end{split}
\end{equation}
Since the current may be expressed in terms of occupation numbers
$\sigma_{11},\ \sigma_{22}$ only we search for an equation equivalent to
\eqref{matrix} for these quantities. Indeed, it is possible to
find such an equation using the normalization condition
$\mathrm{Tr}_S(\rho)=\sum_{i=0}^3\rho_{ii}=1$ and bearing in mind
that $\sigma_{\alpha\alpha}=\rho_{\alpha\alpha}+\rho_{33}$. We may
then write
\begin{equation}
  \begin{pmatrix}
    \Gamma_1 + \mathcal{T}_{12} & -\mathcal{T}_{12} \\
    -\mathcal{T}_{12} & \Gamma_2 + \mathcal{T}_{12}
  \end{pmatrix} \cdot
  \begin{pmatrix} \sigma_{11} \\ \sigma_{22} \end{pmatrix}
  = \begin{pmatrix} \Gamma_{1\uparrow} \\ \Gamma_{2\uparrow} \end{pmatrix}
\end{equation}
with the result
\begin{equation}
  \sigma_{11} - \sigma_{22} = \frac{f_1(\eps_1) - f_2(\eps_2)}
  {1+\mathcal{T}_{12}(\Gamma_1^{-1}+\Gamma_2^{-1})}
  \approx f_1(\eps_1) - f_2(\eps_2) \ .
\end{equation}
We have used the fact that
$\Gamma_{\alpha\uparrow}=f_{\alpha}(\eps_{\alpha})\Gamma_{\alpha}$
and omitted the higher orders terms in $J$ in the last step.

This result together with Eqs.\ \eqref{result},\
\eqref{transmission} predicts the striking phenomenon of
spontaneous particle current between identical baths if
$\eps_1\neq\eps_2$ in the same manner as \C obtained a nonzero
energy flow in \cite{capek}. It is a remarkable fact that the
result does not depend on the nature of the bath, i.e.\ the
permanent current is obtained in analogous manner for the phononic
bath considered by \C as well as for the excitonic (and thus also
fermionic) bath in our case. The only point where the nature of
the bath enters is the formula for the rates and energy
renormalization \eqref{rates}. From the physical point of view
this fact is fully plausible since for a generic bath only its
temperature (and electrochemical potential) should matter in the
thermodynamic predictions.

Now, we come to a contradiction since we know from the exact
solution that there is in fact no such a permanent current. Where
is the problem in the mathematical reasoning and what are its
physical roots? We are going to answer these questions as follows.
First, one should note that the stationary current
$I^{\text{stat}}_1$ is proportional to
$\lambda^6$ (the extra factor $\lambda^2$ comes from
$\Gamma_1+\Gamma_2$). If we really performed the van Hove limit
$t\to\lambda^{-2}t,\ \lambda\to 0$ in which the Davies theory is
valid we would obtain the zero value of the stationary current.

As pointed out by \C in \cite{capek}, when studying a particular
physical system, we cannot actually scale the interaction
strengths to zero since the values are given constants for that
system. At most, we could  have a set of similar physical systems
with different values of coupling constants which could be
considered a series of the same system with the value of
$\lambda$ being gradually scaled down to zero. Then, one expects
that the Davies theory gives more and more precise predictions
about the system's time evolution.

This intuitive formulation has as its mathematical counterpart the
relation
\begin{equation}\label{Davies}
  \lim_{\lambda\to 0}\ \sup_{0\leq \lambda^2 t\leq a}
  ||\rho^\text{exact}(t)-\rho^\text{Davies}(t)||=0 \ ,\text{ for any finite $a$}
\end{equation}
true both for the original and modified Davies theory
\cite{cap-bar-pha-01} which implies an analogous statement for
every matrix component of $\rho(t)$ and, in particular, for
$\text{Im}\rho_{12}(t)$, too. However, this statement is in fact
too weak to be of any practical use for finite $\lambda$'s, as it
is asymptotic by nature.

\begin{figure}
  \centering
  \includegraphics[width=85mm]{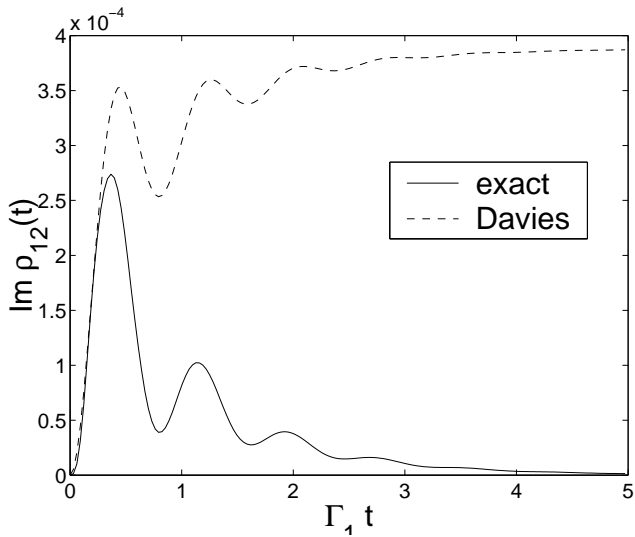}
  \caption{Comparison of the exact evolution of $\text{Im}\rho_{12}(t)$
  with the modified Davies approximation due to \v{C}\'{a}pek. The value of $\lambda$ is
  $1.0$. For the rest of parameters see the text.}\label{obr1}
\end{figure}

\begin{figure}
  \centering
  \includegraphics[width=85mm]{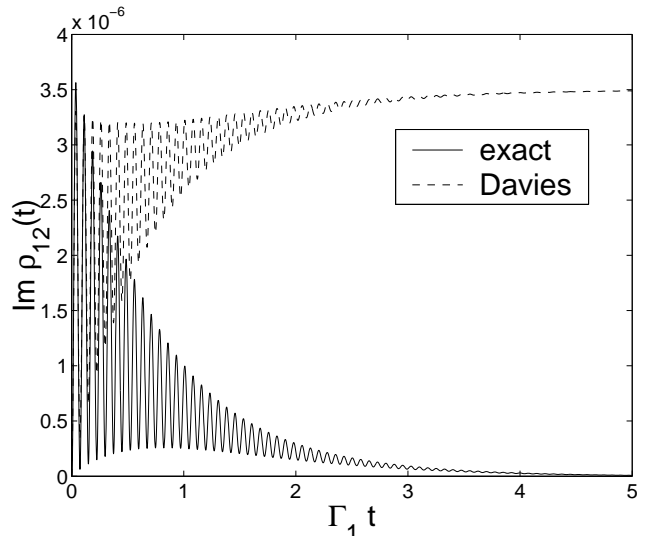}
  \caption{Comparison of the exact evolution of $\text{Im}\rho_{12}(t)$
  with the modified Davies approximation due to \v{C}\'{a}pek for $\lambda=0.3$,
  the other parameters are given in the text.}\label{obr2}
\end{figure}

We may obtain a more precise estimate about the order of the
asymptotic contact if we make an additional assumption (not
implied by the Davies theory) that the $\lambda$ asymptotics is
regular. Then, we can pictorially show how the Davies statement is
in practice realized. In Figs.\ \ref{obr1}, \ref{obr2} there are
depicted
$\text{Im}\rho_{12}(t)$ for the exact and the Davies evolution,
respectively, for our one-dimensional fermionic chain with
$t_0=0,\ \Delta=1,\ T_1=T_2=0.1,\ \mu_1=\mu_2=0,\ \epsilon_1=-0.1,\ \epsilon_2=0.2,
\ J=0.001,\ V_1=0.1,\ V_2=0.15,\ \rho_{00}(t_0)=1 \text{
(initially empty dot)}$ for $\lambda=1\text{ and }0.3$.

\begin{figure}
  \centering
  \includegraphics[width=85mm]{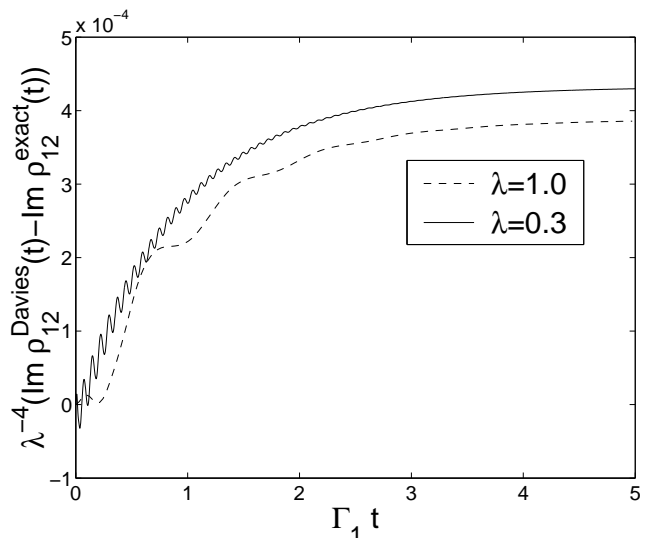}
  \caption{The difference between the approximate and exact evolution of the element
  $\text{Im}\rho_{12}(t)$ for two values of $\lambda$.}\label{obr3}
\end{figure}

In Fig.\ \ref{obr3} there is shown the scaled difference
$\lambda^{-4}\text{Im}(\rho^\text{Davies}_{12}(t)-\rho^\text{exact}_{12}(t))$
for the two $\lambda$'s. This clearly shows that the difference
between these two quantities for any finite scaled time goes to
zero roughly as $\lambda^4$ in agreement with the Davies
statement. Yet the exact result does not exhibit any permanent
stationary current. Indeed, when one thoroughly inspects the
Davies formula \eqref{Davies} one has to come to the result that
it is fully consistent with the zero stationary current as
illustrated by our pictures. On the other hand it does not exclude
the possibility of a nonzero value in general which only means
that its predictive power concerning this issue is essentially
zero. The conclusion drawn by \C from it is doubtful since what he
finds to be the stationary current breaking the second law of
thermodynamics is of the same order in $\lambda$ as the terms
neglected in a systematic Davies theory.

Now, let us discuss the physical mechanism of the above apparent
paradox and the role of the quantum mechanics in it. As obvious
from the above, the Davies theory neglects the higher order
processes in $\lambda$. Actually, it can be considered as a sort
of quasi-classical limit yielding the Pauli equation while
omitting higher order quantum mechanical processes. Obviously, it
does not take into account properly processes of direct coherent
tunneling between the baths for finite $\lambda$. Most probably,
just these higher order coherent processes exactly cancel the
spurious stationary quasi-classical current as one can infer by a
comparison of the exact transport formula \eqref{exact} with the
Davies one \eqref{result}. Thus, referring to the conjectures
mentioned in the introduction, for this particular model the
coherent quantum mechanical features of the model prevent the
second law from being violated rather than allowing it. It is
interesting that an analogous discrepancy between the two
approaches (reduced density matrix versus NGF) was reported by
Wacker \cite{wacker} in a more complicated transport study.

To conclude, we have presented an exactly solvable model of
quantum transport and used it to test the validity of the
predictions by \C about the violation of the second thermodynamics
law. We found, however, these predictions based on the Davies
theory, rigorous in itself, as unwarranted. The point is that the
predicted permanent current (or energy flow) is within the error
of the asymptotic Davies theory for any finite coupling strength.

\begin{acknowledgments}
This work is a part of the research program MSM113200002 that is
financed by the Ministry of Education of the Czech Republic.
Support of the grant 202/01/D099 of the Czech grant agency is also
gratefully acknowledged.
\end{acknowledgments}


\end{document}